\documentclass[runningheads]{llncs}
\usepackage{amsmath}
\usepackage{amssymb}
\usepackage{graphicx}
\usepackage{graphics}
\usepackage{listings}
\usepackage{multirow}

\newtheorem{fact}{Fact}

\title{Geometric approach to string analysis: \\ deviation from linearity and its use\\ for biosequence classification}

\titlerunning{Geometric approach to string analysis}

\author{Boris Brimkov\inst{1} \and Valentin E. Brimkov\inst{2}}

\authorrunning{B. Brimkov and V. Brimkov}

\institute{
Computational \& Applied Mathematics, Rice University, Houston, TX 77005, USA\\
\email{boris.brimkov@rice.edu}
\and Mathematics Department, SUNY Buffalo State College, Buffalo, NY 14222, USA\\
\email{brimkove@buffalostate.edu} 
}

\begin{document}
\maketitle

\begin{abstract}
Tools that effectively analyze and compare sequences are of great importance in various areas of applied computational research, especially in the framework of molecular biology. 
In the present paper, we introduce simple geometric criteria based on the notion of \emph{string linearity} and use them to compare DNA sequences of various organisms, as well as to distinguish them from random sequences.
Our experiments reveal a significant difference between biosequences and random sequences –-- the former having much higher deviation from linearity  than the latter --– as well as a general trend of increasing deviation from linearity between primitive and biologically complex organisms.  

\medskip

\end{abstract}

\section{Introduction}
The theory of words studies the structural properties of strings composed from letters of a given alphabet, and provides algorithms for solving diverse problems defined on strings. Among the most important motivations of the discipline is its relevance to computational biology, and more precisely, to the automated analysis of biosequences. This includes a great variety of problems whose portrayal is beyond the purposes of the present paper. Some avenues of the ongoing research are surveyed in \cite{sankoff,waterman,axa0}. In particular, an important task is identifying certain patterns, motifs, or biologically meaningful features in a given biosequence.

Typically, the considered problems are approached using combinatorial techniques such as combinatorial pattern matching and combinatorics on words. In this paper we instead use a \emph{geometric} approach in an attempt to address questions that are important for understanding biological evolution.    

A number of past studies have attempted to address  by quantitative means the question of what distinguishes biosequences from random sequences. While by its very nature such a goal has been found quite elusive \cite{broox}, there is substantial evidence in support of the argument that biosequences feature properties that are typical of random sequences (for example, near-total incompressibility \cite{nevil}). Thus, biosequences are regarded as ``slightly edited random sequences" \cite{weiss}, and modern proteins are believed to be ``memorized" ancestral random polypeptides which have been slightly modified by the evolutionary selection process in order to optimize their stability under specific physiological conditions \cite{axa2}. Biosequences appear to be hardly distinguishable from their random permutations, although the latter are clearly incongruous with living organisms \cite{monod,schwartzKing,whiteJacobs}. While this may seem quite obvious from a biological point of view, there have also been numerous computational arguments that support this claim. For example, in \cite{pande} Pande et al. present results of mapping some protein sequences onto so-called Brownian bridges, which revealed a certain deviation from randomness. In another study, by estimating the differential entropy and context-free grammar complexity, Weiss et al. have shown that the complexity of large sets of non-homologous proteins is lower than the complexity of the corresponding sets of random strings by approximately 1 \% \cite{weiss}.
As a first major result of the present work, we introduce simple geometric criteria by which biosequences very strongly differ from random sequences of the same length.
In view of the above-mentioned 1\% difference demonstrated in \cite{weiss}, by ``very strongly" we refer to differences in the order of several hundred percent, registered for 25 biosequences compared to random sequences over the same alphabet and length.

Furthermore, provided the widely adopted postulates of the theory of evolution and in view of the available theoretical and experimental results, it is natural to conjecture that in the evolutionary process of organisms from primitive to biologically complex, their corresponding biosequences have been evolving from random or close to random toward ones that feature increasing deviation from randomness. 
As a second major result, our experiments based on the introduced measures confirm this expectation (although not in equally indisputable terms as for the comparison between random sequences and biosequences). That is also in accordance with results suggesting that biosequences of proteins which are close in the genome are more similar than those of proteins far apart in the genome. 


To this end, we use a discrete geometric approach. Given a string $s=s_0 \hdots s_m$ over an alphabet $X$ with $|X|=n$, we define an ordered set $L(s)=p_0 \hdots p_m$ of points which form a discrete monotone path in $\mathbb{Z}^n$.
We then define the deviation of such a monotone path from linearity, and state some basic properties related to this notion. 
 
After introducing several deviation measures, we apply them to the biosequences of a set of organisms that stand at different levels of the evolution scale (i.e., from primitive organisms such as microbes, through plants and reptiles, up to mammals). We also compare all these with random sequences of the same length.
The obtained results demonstrate a significant difference between biosequences and random sequences -- the former being much further from linearity than the latter -- as well as a general trend of increasing deviation from linearity between primitive and biologically complex organisms. Results of some other related experiments are outlined as well. 

The paper is organized as follows. In the next section we introduce some technical notions and notations, including ones from the theory of words. In Section~3 we introduce the notions of string linearity and deviation from linearity, and study several related properties. In Sections 4 and 5, we present our experimental results and offer a short discussion. We conclude with final remarks and open questions in Section 6.

\section{Definitions and notations} 

\subsection{General}

By $|X|$ we denote the cardinality of set $X$ and by $\overline{xy}$ the straight line segment with endpoints $x$ and $y$. By $d(x,y)$ we denote the Euclidean distance between points $x$ and $y$, and by $d(x,Y)$ the distance between point $x$ and set $Y$, i.e., 

$$d(x,Y)=\inf_{y \in Y}\{d(x,y)\}.$$

Given a list $T$ of nonnegative real numbers $t_1 \hdots t_k$ (not all of which equal 0), a \emph{normalization} of $T$ is obtained by multiplying each value in $T$ by $\frac{100}{t_{\max}}$ where $t_{\max}= \max_{1 \leq i \leq k} \{t_i\}.$

Given an approximation algorithm $A$ for a minimization problem $\Pi$ with a set of instances $D_{\Pi}$, let $A(I)$ be the value of an approximate solution to instance $I$ found by $A$. 
The \emph{approximation ratio} of $A$ on $I$ is $R_A(I)=\frac{A(I)}{Opt(I)}$, where $Opt(I)$ is the optimal solution for $I$; the \emph{worst case performance ratio} of $A$ is $R_A=\sup \{ R_A(I) :  I \in D_{\Pi} \}$. 
We will say that an algorithm with performance ratio $r$ finds an $r$-\emph{approximation} to the optimal solution.
For more details the reader is referred to \cite{cormen,GJ}.

\subsection{Notions of theory of words}

In the literature, the terms word, sequence, and string are often used interchangeably. A \emph{sequence} is often defined in mathematics as a function whose domain consists of a set of consecutive integers, and a \emph{string over X}, where $X$ is a finite set, is often defined as a finite sequence $s$ of elements from $X$ ($X$ is also sometimes called the \emph{alphabet}). The term \emph{word} is frequently used as an abstraction of the other two terms. In biology, the prevalent term is \emph{biosequence}; biosequences are built from the four letters A, T, C, G, and have finite length. 

The theory of words is a central topic in theoretical computer science \cite{crochemore,axaGalil}. Below, we recall a few basic notions and fix some denotations to be used in this paper.

In string $s=s_0 \hdots s_m$ over set $X$, $s_i$ is the $i^{\mathrm{th}}$ term of $s$ ($0 \leq i \leq m$), which is some element of $X$. The number of elements in $s$ is called the \emph{length} of $s$ and denoted $|s|$. If $|s|=0$, we say that $s$ is the \emph{empty string}, denoted by $\lambda$. We denote $k \geq 1$ consecutive repetitions of term $x$ in string $s$ by $x^k$.

If $s$ and $t$ are two strings, the string consisting of $s$ followed by $t$, written $st$, is called the $concatenation$ of $s$ and $t$. A $substring$ of a string $s$ is obtained by selecting some or all consecutive elements of $s$. More formally, a string $v$ is a substring of the string $s$ if there are strings $u$ and $w$ such that $s=uvw$ (where we may have $u=\lambda$ or $v=\lambda$ or $u=v=\lambda$).

Following the terminology introduced in \cite{axa1,axa2}, given a string $s$, a {\em subsequence} of $s$ is any string $u$ which can be obtained by removing from $s$ one or more, not necessarily consecutive terms.

\section{String geometrization}

To present our approach, we conform to a digital geometry setting in which the considerations take place in the \emph{grid cell model}.  
In this model, the regular orthogonal grid defines a partition of $\mathbb{R}^n$ into
$n$-dimensional hypercubes (e.g., unit squares for $n=2$, unit cubes 
for $n=3$) also called \emph{n-cells} or \emph{voxels}.
The $n$-cells are centered at the grid points and their edges are parallel to the coordinate axes. See \cite{klette} for more details.

Let $s = s_0 \hdots s_m$ be a string on an alphabet $X = \{x_1, \dots,x_n\}$. We inductively construct an ordered set $L(s)$ of points $p_0, \hdots, p_m$ corresponding to string $s$ as follows.

We set $p_0$ to be the origin of the Cartesian coordinate system. Let $p_i=(p_{i,1},\hdots,p_{i,n})$ be the $i^{\mathrm{th}}$ element of $L$ for $0 \leq i <m$. 
If $s_{i+1} = x_j$  for some $j$ $1 \leq j \leq n$, then we set $p_{i+1}=(p_{i,1}, \hdots, p_{i,j}+1, \hdots, p_{i,n})$.
Thus, we obtain a {\em monotone} discrete path $L(s)$ with $|L(s)|=|s|=m+1$, in which the coordinates of a point are pairwise greater than or equal to the corresponding coordinates of any preceding point. 
Fig. 1, left, gives an example of a string over the alphabet $\{x,y\}$ and its corresponding monotone discrete path.

\begin{figure}
\begin{center}
\includegraphics[scale=0.3]{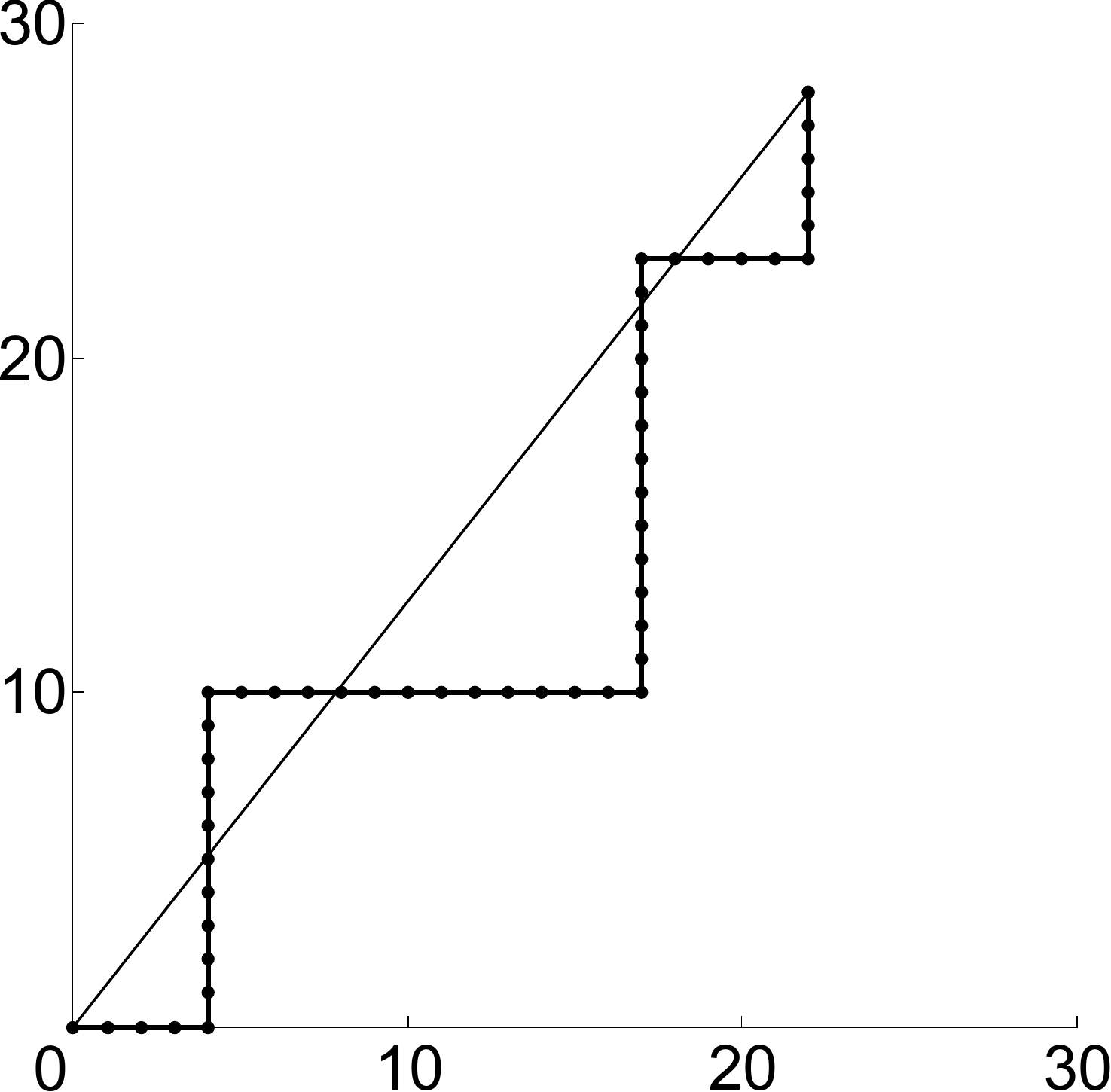} \quad \qquad \qquad
\includegraphics[scale=0.3]{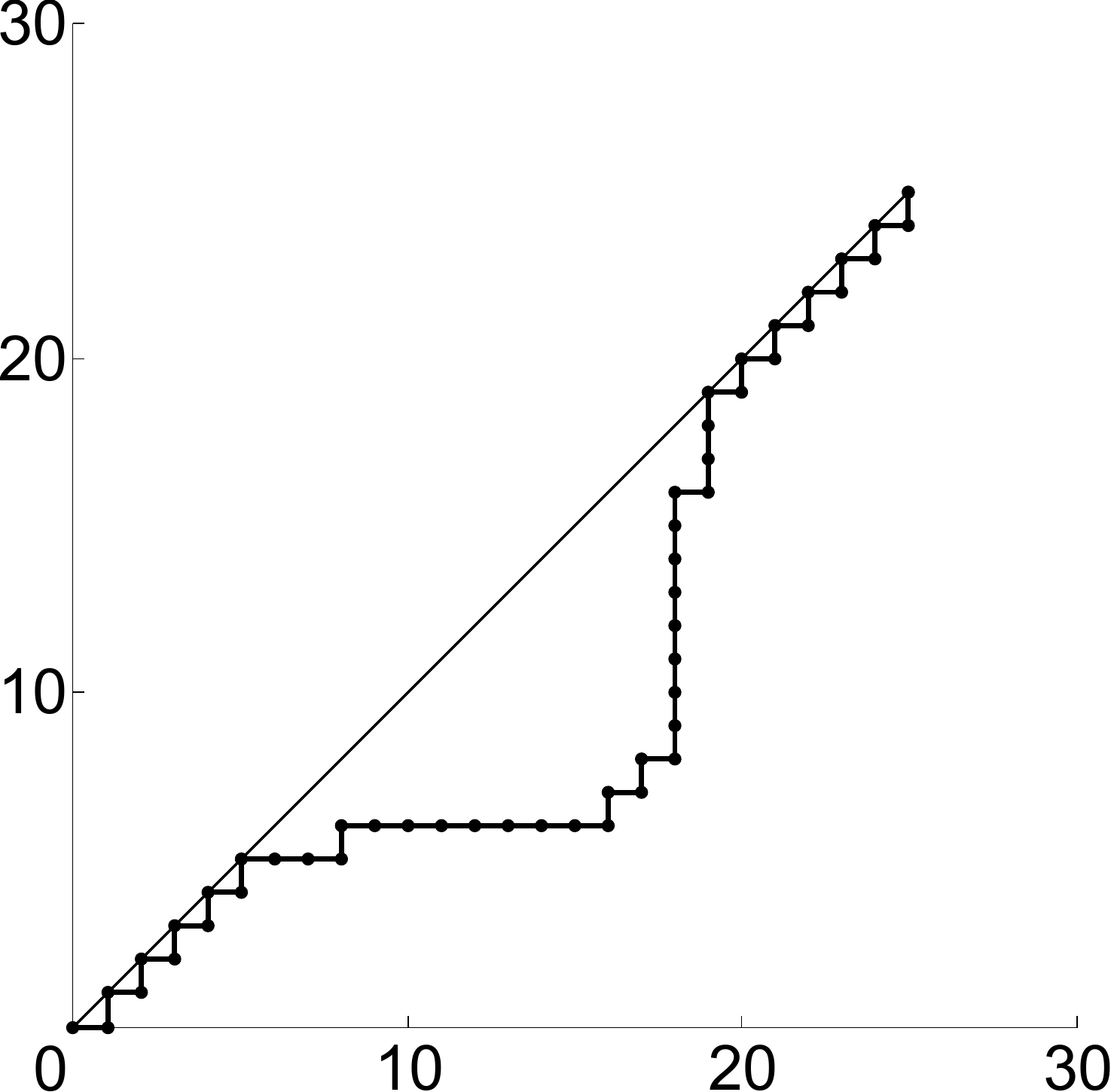}
\end{center}
\caption{Left: the string $s=x^4y^{10}x^{13}y^{13}x^5y^5$ and its corresponding monotone discrete path. Right: another string which has nearly the same $adv$ and $mdv$ as $s$, but a different $nlm$.}
\end{figure}

Having  such a discrete path constructed, one can study its geometric and combinatorial properties, which in turn can provide useful information about the original string $s$. 
Clearly, the properties and characteristics of monotone discrete paths, not necessarily representing strings, could be interesting in their own right.

Let $p_0$ be the origin and $p=(p_1,\hdots,p_n)$ be a point in $n$-dimensional space $\mathbb{Z}^n$. 
Denote by $\mathbb{H}$ the set of all monotone discrete paths between $p_0$ and $p$. It is easy to see that 
$$|\mathbb{H}| = \frac{(p_1+\hdots+p_n)!}{p_1!\hdots p_n!}.$$ 
Each path $H \in \mathbb{H}$ consists of $1+\sum_{i=1}^n p_i$ points, with initial point $p_0$ and terminal point $p$. 
If for every point $h$ in $H$, the voxel centered around $h$ intersects the line segment $\overline{p_0p}$, we call $H$ a \emph{linear path}. 
Accordingly, we call a string \emph{linear} if its corresponding monotone path is linear.

It is easy to see that the following facts hold:

\begin{fact}
Given a line segment $\overline{p_0p}$, there is at least one linear path from $p_0$ to $p$.
\end{fact}  

\begin{fact}
If $H$ is a linear path from $p_0$ to $p$, then $d(h,\overline{p_0p}) \leq \frac{\sqrt{n}}{2}$ $\forall h \in H$.
\end{fact}  

Next we define some string characteristics that are instrumental to the experimental studies presented in the subsequent sections.

Let $s$ be a string and $L(s)= p_0 \hdots p_m$ be its corresponding monotone path. 
We define the \emph{maximum deviation of $s$ from linearity} as 
$$
mdv(s) = \max_{i=0}^m \{d(p_i,\overline{p_0p_m})\},
$$ 
and {\em average deviation of $s$ from linearity} as 
$$
adv(s) = \frac{\sum_{i=0}^m d(p_i,\overline{p_0p_m})}{m+1}.
$$ 

\begin{remark}
\label{r1}
Note that when $n=4$ (which is the case for biosequences), the $adv$ and $mdv$ of a linear string are at most 1.
\end{remark}

The third characteristic of a string $s$ will be called the {\em number of local maxima} of $s$ and denoted $nlm(s)$.  
Formally, by a local maximum we mean a point $p_i \in L$ for which 
$d(p_i,\overline{p_0p_m})$ is greater than $d(p_{i-1},\overline{p_0p_m})$ and $d(p_{i+1},\overline{p_0p_m})$.
However, regarding the usual applications of extrema of discrete functions, in particular in view of our own purposes, counting all such maxima does not seem to be very relevant. 
Instead, local maxima can be counted only if they ``stand out" compared to other, ``indistinguishable" local extrema, which differ very little from neighboring points. 
Thus, we adopt the notion of number of local maxima as ``method dependent." 
Specifically, our choice of method is the one provided by \cite{yoder}. 
The $nlm$ measure may be useful to distinguish between strings which have the same $adv$ and $mdv$, as illustrated in Fig. 1, right.


\subsection{Relation to minimum enclosing cylinder} 

Our definition of a linear string refers to a discrete monotone path whose voxels intersect the line segment between the initial and terminal points.
Respectively, deviation from linearity refers to the distances from the points of the monotone path to the line segment.

Another possible approach is to consider a straight line that minimizes the maximal distance over all points of $L(s)$, which is the axis of the minimum enclosing cylinder for the set $L(s)$. 
Note that while in two dimensions the problem of finding the minimum enclosing cylinder can efficiently be solved in linear time, it is not so in higher dimensions.
Even in 3D, the available exact algorithms take super-cubic time (see, e.g., \cite{agarwal,chan,preparata}). 
This  makes the problem practically intractable  for strings of considerable size, e.g., like the biosequences we investigate in the following sections.  
The following proposition demonstrates that the deviation from linearity which we adopt is no more than twice greater than the one defined by the minimum enclosing cylinder.  
Moreover, the computation of the former requires only a linear number of operations for a fixed dimension $n$ (4 in the case of biosequences), and is therefore without a doubt advantageous from a computational complexity perspective.
Note also that in the course of our experiments, we measured a significant difference between the deviation from linearity of biosequences and random sequences; thus, a minimum enclosing cylinder approach -- provided that one could afford to wait for the solution -- would provide no advantage in distinguishing biosequences from random sequences aside from changing the magnitude of distinction by at most a factor of 2.

\begin{proposition} 
Let $L=p_0 \hdots p_m \subset \mathbb{Z}^n$ be a monotone discrete path of length at least 3.
A 2-approximation to a minimum enclosing cylinder for $L$ can be found with $O(mn)$ operations. 
\label{prop} 
\end{proposition}
{\bf Proof} \
Let $l$ be the straight line through the first and the last points of path $L$,
which is the axis of an enclosing cylinder for $L$.
Let $p$ be a point of $L$ which maximizes the distance to $l$ and denote $r = d(p,l)$.  

Let $C^*$ be an enclosing cylinder for $L$ of minimal radius $r^*$ centered about an axis $l^*$.
We have $r^* \leq r \leq 2r^*$, as the second inequality holds form the following argument.
Let $p' \in l$ be the foot of the perpendicular from $p$ to $l$, i.e., $d(p,p')=r$.
By the construction of path $L$ and plain geometric arguments, 
we have that a minimal enclosing cylinder $\bar C$ for the three points  $p_0, p_m$, and $p$ has axis $\bar l$ that is parallel to line $l$
and passes through the midpoint of segment $\overline{pp'}$; obviously, the radius $\bar r$ of $\bar C$ satisfies $\bar r = r/2$.
Since $\{ p_0,p_m,p \} \subseteq L$, we have $\bar r \leq r^*$ and thus $r \leq 2 r^*$.

\medskip

The estimate of the time necessary to compute $r$ follows from related calculus formulas.  
Let $l$ have a parametric equation $x = t a$, where $x,a \in \mathbb{R}^n$, $a=(a_1,\dots,a_n)$, 
and let point $p'$ be as defined above.
$p'$ belongs to $l$, so $p'=t'a$ for some $t' \in R$.
Vector $p'-p$ is orthogonal to vector $a$, i.e., their scalar product satisfies $a \cdot (p'-p)=0$. 
From the last equality one easily obtains
\begin{equation}
t'=\frac{a_1p_1+\dots+a_np_n}{a_1^2+\dots+a_n^2} \ \ \ {\rm and} \ \ \ d(p,l) = \sqrt{(a_1t'-p_1)^2 + \dots + (a_nt'-p_n)^2}
\end{equation}

Obviously, $t'$ and $r=d(p,l)$ can be found with $O(n)$ arithmetic operations and a single square root computation; the latter is not necessary to perform when comparing distances from points of $L$ to line $l$.  
\qed

\section{Deviation from linearity of random sequences and biosequences: experimental study}

\subsection{General description of experimental procedures}

The notion of string linearity furnishes an easily implementable tool to compare the biosequences of various organisms. It is reasonable to conjecture that biologically complex organisms have highly structured DNA whose corresponding monotone path strongly deviates from a straight line, while primitive organisms have less structured DNA, whose corresponding monotone path is closer to a straight line. Moreover, a completely random sequence over the alphabet $\{A, T, C, G\}$ has no structure, and therefore its corresponding monotone path can be expected to be much closer to a straight line. 

To test this hypothesis, we compare the deviation from linearity of the biosequences of 25 organisms with varying biological complexity, as well as that of random sequences. The number and type of organisms we consider is typical of comparative analysis studies in molecular biology (see, e.g., \cite{salzburger}). We took the biosequences from the genome-scale repository and browser Ensembl Genomes, which is managed by the European Bioinformatics Institute. For each organism, we processed relatively short substrings and subsequences of DNA in FASTA format, selected randomly from an excerpt of the genome containing about two million letters. Ideally, it might be more informative to select samples from larger excerpts of the genomes, or even to process the entire genomes of organisms at once, since the (absolute) deviation from linearity of a string is generally dependent on its length.

In our experimental study we first studied the effects of string size on the proposed linearity measures and then selected a suitable string size for our more extensive experiments. This helped save computation time and in turn allowed us to repeat each procedure 1,000 times with different samples of the studied genomes, thus increasing the confidence in the obtained results. Note that some organisms' genomes have billions of letters (whose processing would require a lot of time), while others have genomes that are still uncharted or studied only partially.

Thus, we operated under the assumption that a comparison of the linearity of samples of the genomes of a certain reasonable size, that is ``comparatively small" but ``sufficiently large", will classify organisms in the same \emph{relative} order as a comparison of the linearity of their whole genomes. As we will see in the following section, our experiments support this claim.

Recall also that we distinguish between substrings and subsequences of a given string $s$, the former being segments of consecutive terms while the latter being ordered subsets of not necessarily consecutive terms of $s$.  Typically, experimental research involving biosequences is based on processing families of substrings rather than subsequences. Only recently, Apostolico and Cunial attempted to assess (``perhaps for the first time," as these authors believe) the structure and randomness of polypeptides in terms of subsequences satisfying certain conditions \cite{axa1,axa2}. As the results obtained therein seem interesting and promising, we performed all our experiments both on substrings and subsequences. The similarities and differences within both frameworks are presented and discussed in the following sections. 

\subsection{Effects of substring and subsequence length}

We investigated how the length of a biosequence affects its deviation from linearity, and ascertained that deviation increases with the size of the string. However, the rate of increase seems to be independent from the \emph{type} of organism, and therefore an organism's deviation from linearity \emph{relative} to the deviation of other organisms is independent of the length of the biosequences, as long as the length is constant across organisms. These claims are supported by Figure~2, which shows the absolute and normalized maximum and average deviations from linearity of different organisms measured for substrings and subsequences of increasing length.

The top four graphs display the $adv$ and $mdv$ computed for substrings, and the bottom four graphs display the $adv$ and $mdv$ computed for subsequences. The left four graphs display the absolute $adv$ and $mdv$ and the right four graphs display the normalized $adv$ and $mdv$ for the graphs on their left. Note that the graphs in the two columns are essentially the same, where the normalized graphs in the right column are the result of ``pulling up" the left sides of the graphs in the left column.

\begin{figure}
\begin{center}
\includegraphics[scale=0.185]{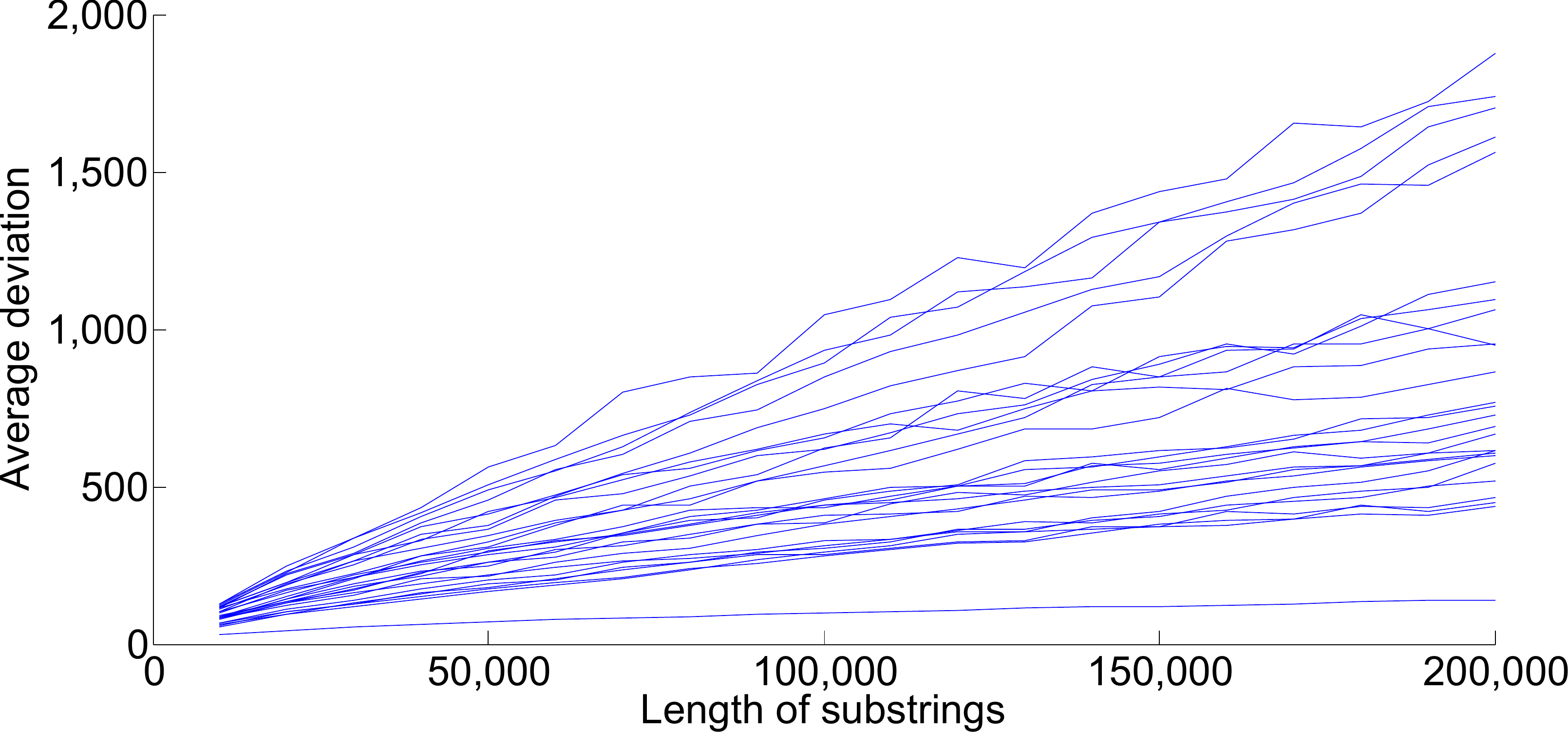}
\includegraphics[scale=0.185]{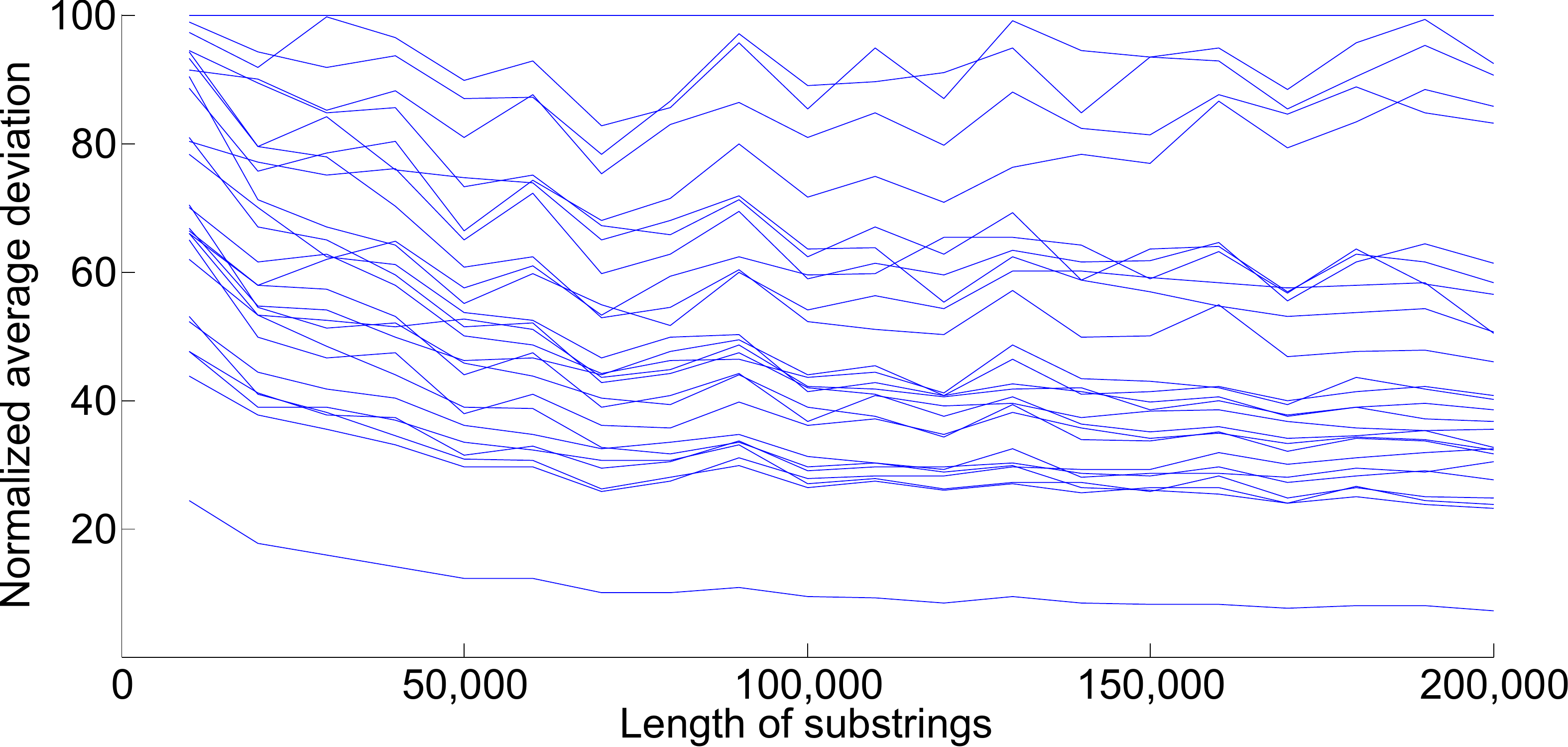}
\includegraphics[scale=0.185]{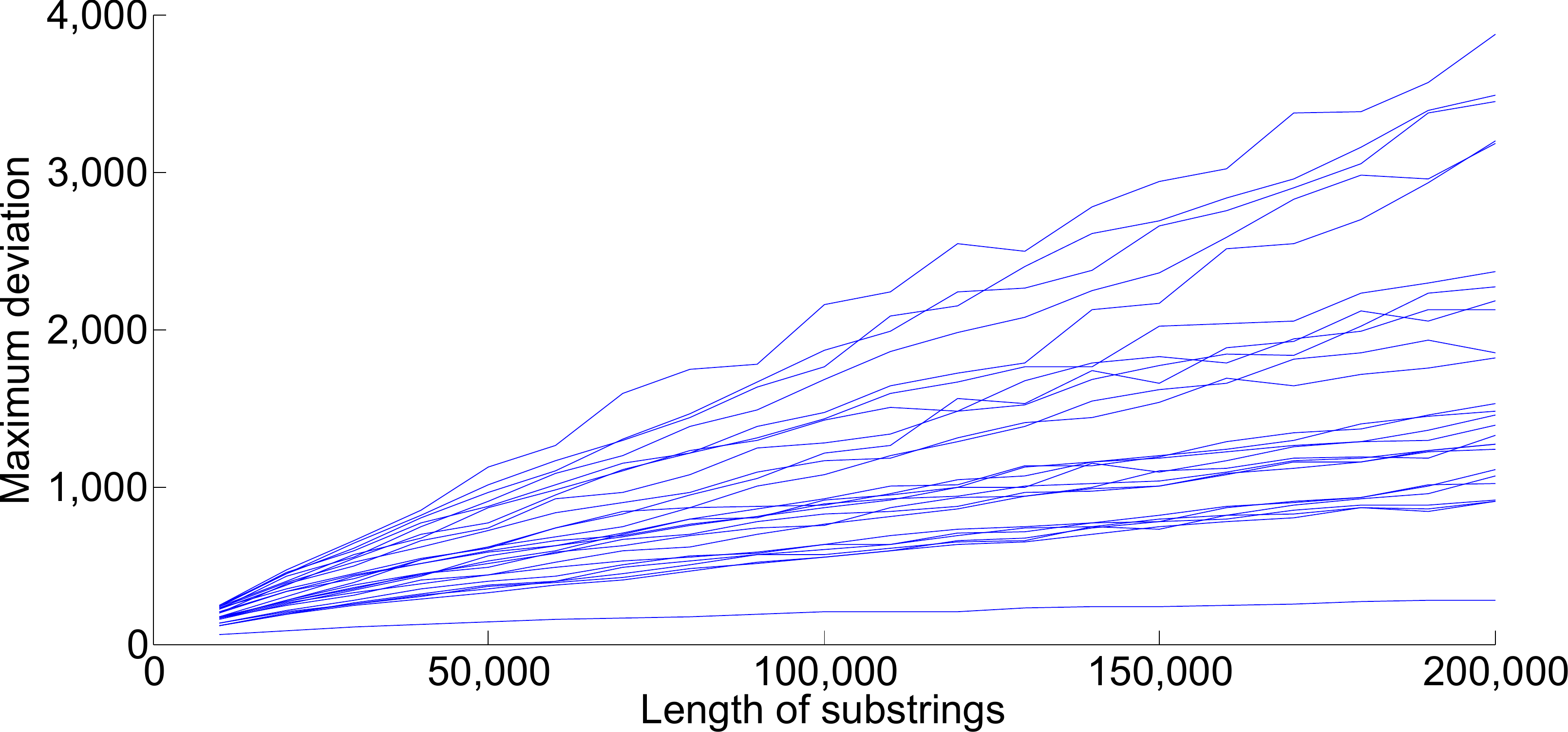}
\includegraphics[scale=0.185]{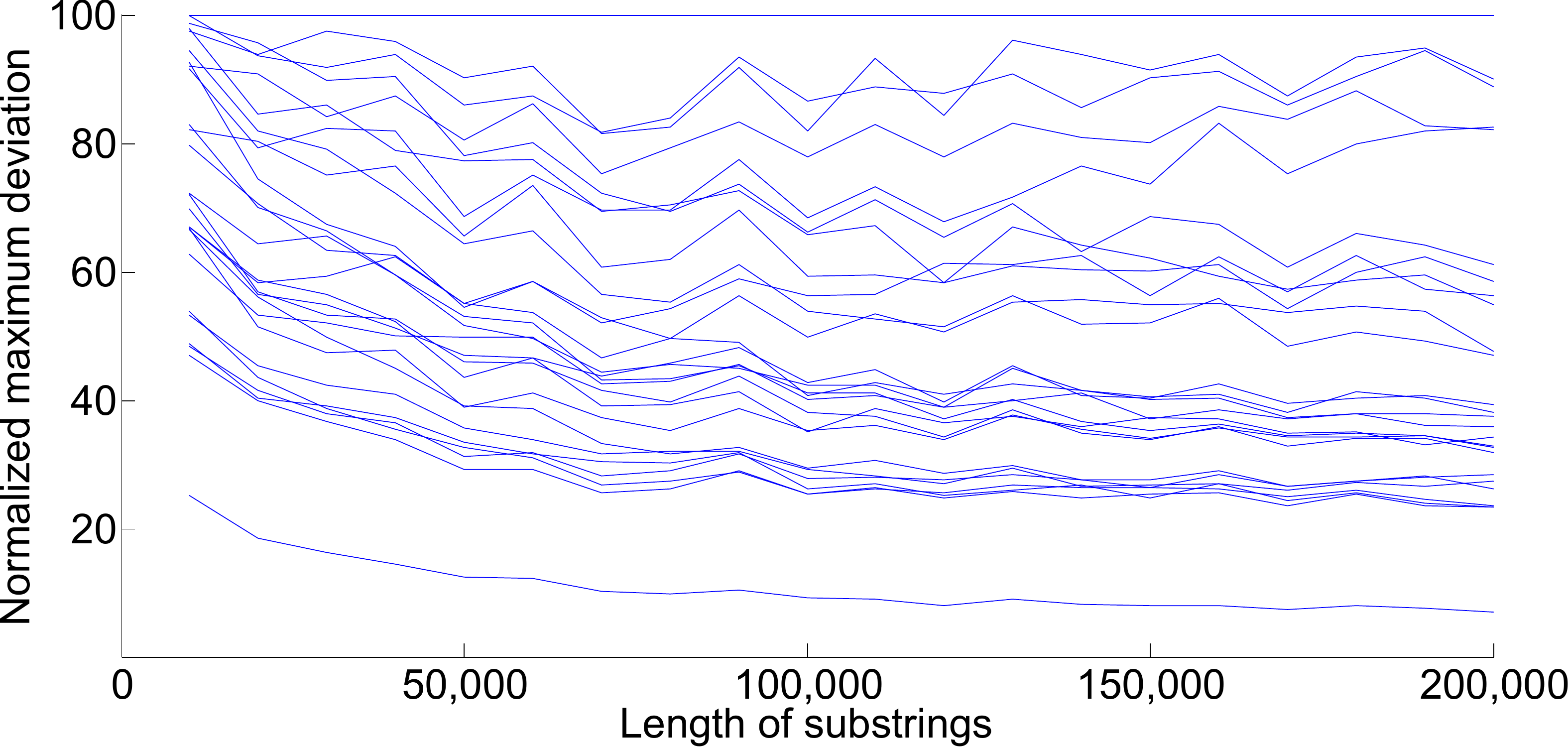}
\includegraphics[scale=0.185]{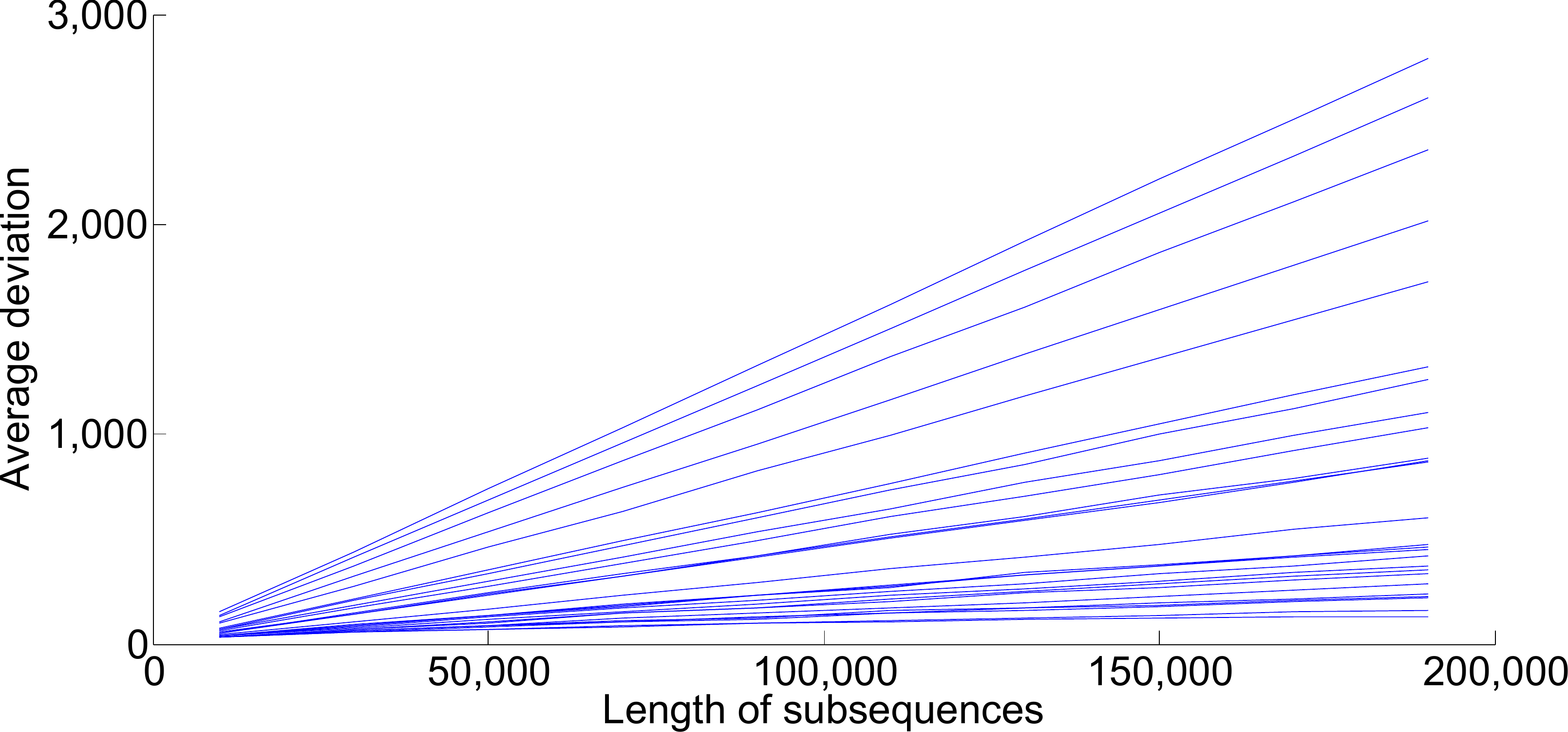}
\includegraphics[scale=0.185]{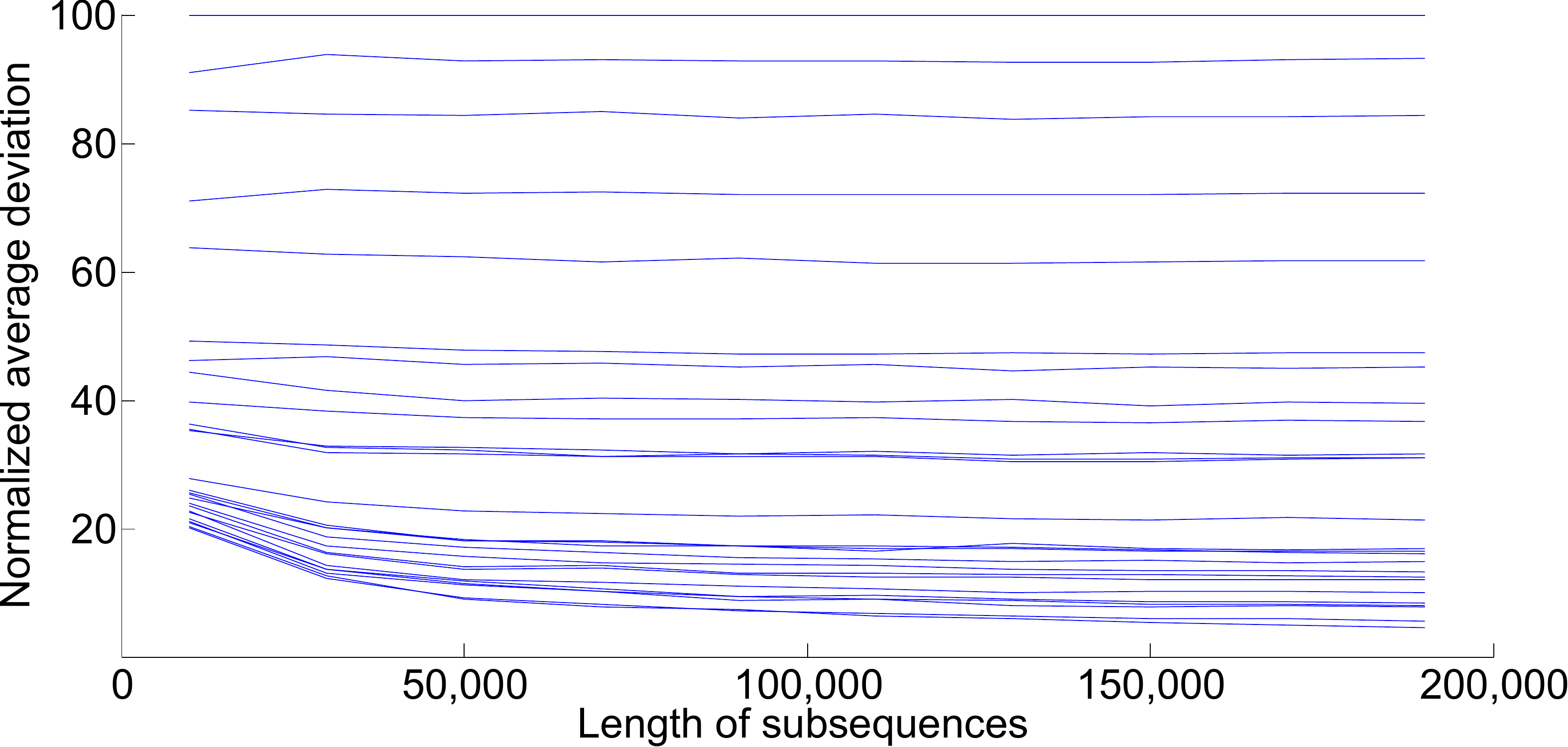}
\includegraphics[scale=0.185]{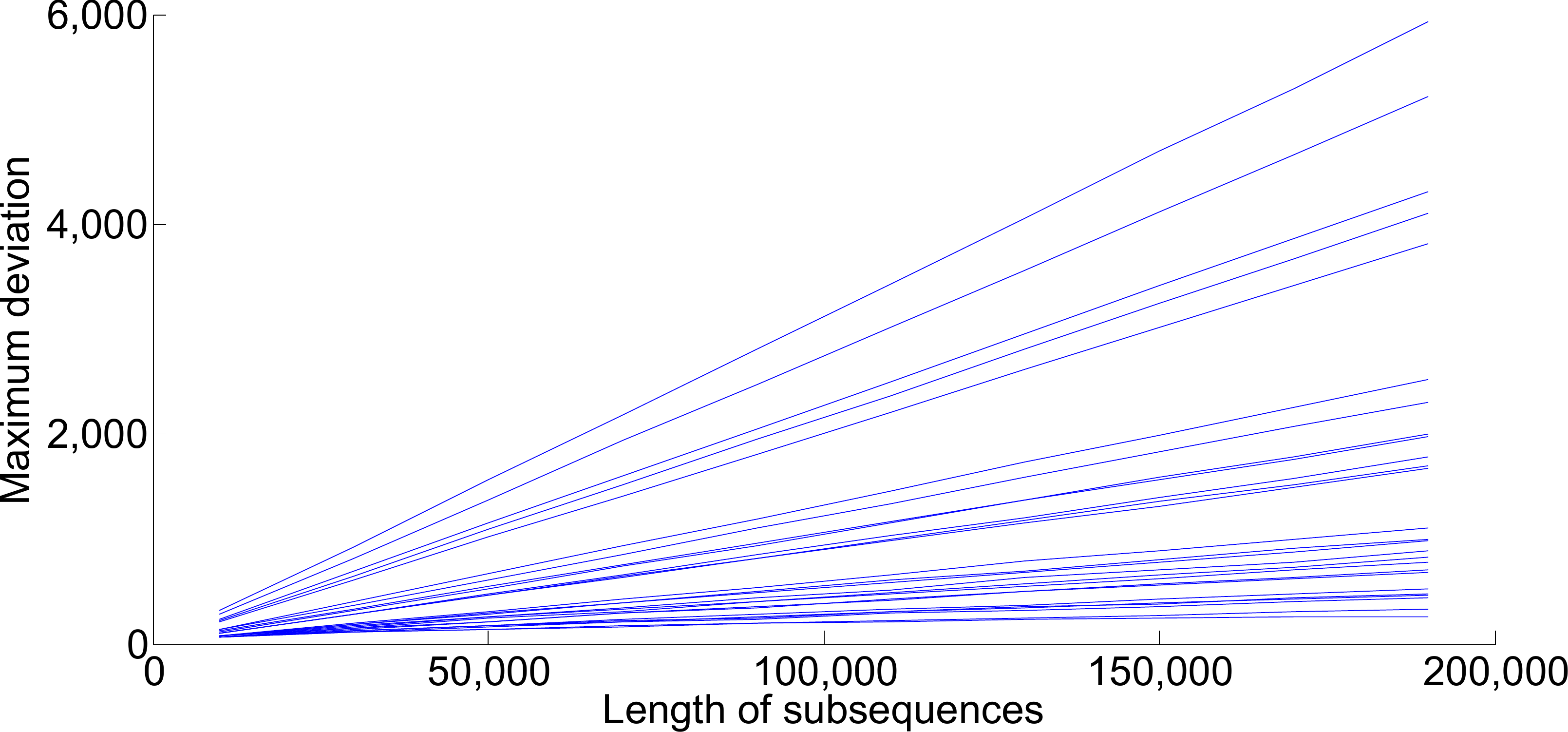}
\includegraphics[scale=0.185]{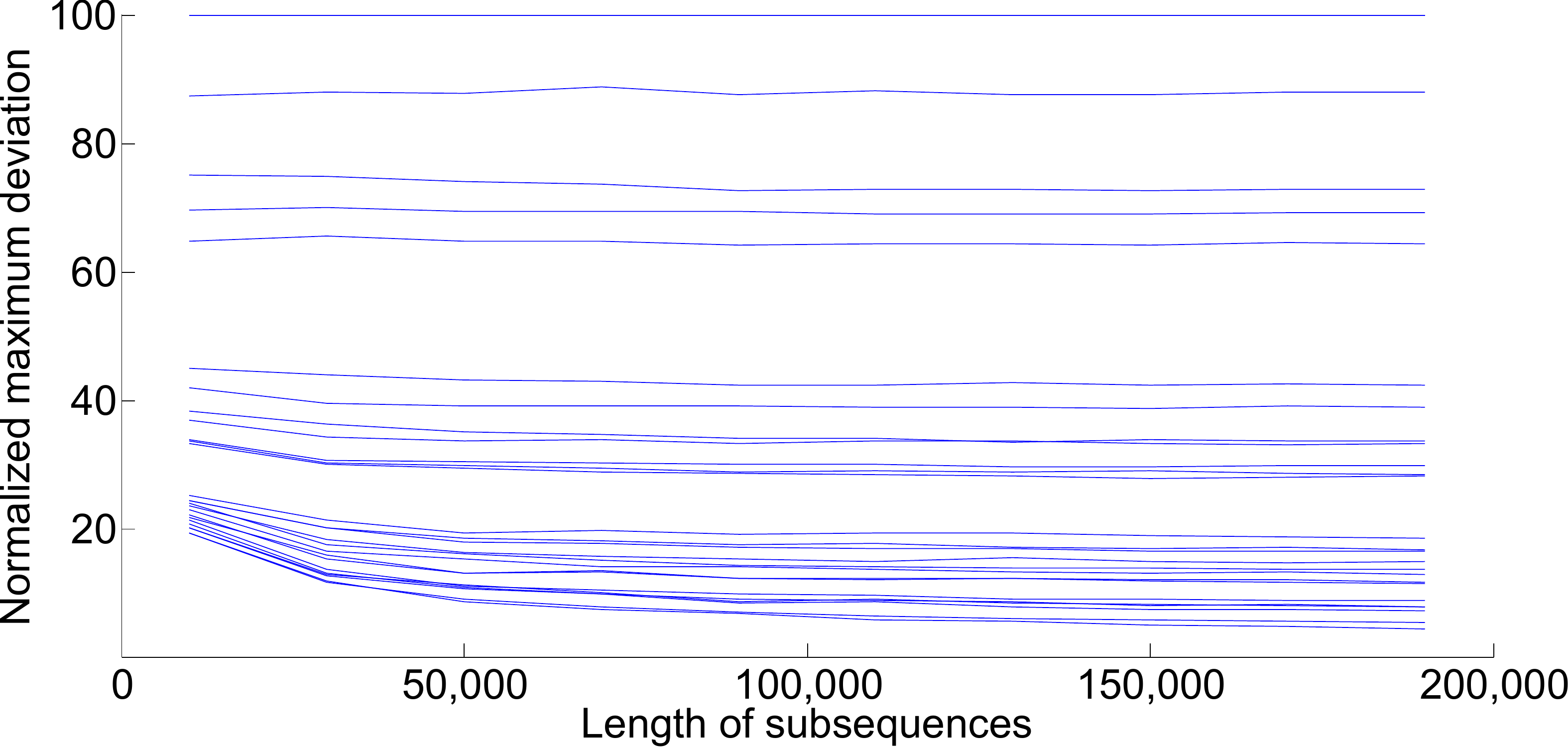}
\caption{Absolute and normalized $avd$ and $mdv$ measured for substrings and subsequences of increasing length.}
\end{center}
\end{figure}

For all graphs, the substrings and subsequences are taken randomly from the 26 sources listed in Table 1. For the top four graphs, the corresponding linearity measures were computed for substrings of length $10,000 \times k$ for $1 \leq k \leq 20$; for each of these lengths, the linearity measures were computed for 200 different substrings taken randomly from each of the 26 sources, and the average values of the 200 trials were plotted. For the bottom four graphs, the corresponding linearity measures were computed for subsequences of length $10,000 \times (2k-1)$ for $1 \leq k \leq 10$; for each of these lengths, the linearity measures were computed for 100 different subsequences taken randomly for each of the 26 sources, and the average values of the 100 trials were plotted.\footnote{Fewer lengths and trials were used for subsequences because adding more ``resolution" to the almost completely straight lines in these graphs does not gain us any useful information.}

From Figure 2, it can be seen that $adv$ and $mdv$ measured from subsequences are more independent of length than $adv$ and $mdv$ measured from substrings, since there is less crossing between lines in the bottom four graphs. However, even when measured from substrings, $adv$ and $mdv$ are principally independent of length, since for any of the lengths examined the organisms are more or less in the same relative position compared to the other organisms.

Clearly, as the length of substrings and subsequences approaches 0, the measures of deviation from linearity will approach 0, and will be more unstable and unreliable. From the normalized graphs, we notice that the initial fluctuations first disappear around substrings and subsequences of length 50,000. For this reason, we carried out our further experiments (which involve more trials and hence a greater confidence) with substrings and subsequences of this length.

\subsection{Description of computational procedure}

We used version R2011a of Matlab to carry out our computations. The built-in Matlab functions \texttt{randi} and \texttt{randseq} were applied to make random selections; additionally, we used a function called \texttt{peakfinder}, provided by \cite{yoder}, to find local extrema. 
Our computational process can be broken up into the following components: 

\begin{enumerate}
\item Selection of samples
\item Computation of linearity measures
\item Compilation and normalization of data
\end{enumerate}

\subsubsection{Selection of samples}
To select a random substring $b_1 \hdots b_m$ from a larger string $a_1 \hdots a_n$, we pick a random integer $i$ from the interval $[1,n-m+1]$ using Matlab's \texttt{randi} function; then, we set $b_j=a_{i+j-1}$ for $1 \leq j \leq m$. In this manner, we select substrings from each of the organisms and store the substrings, which all have length $m$, in an array. We also store a random sequence of length $m$ over the alphabet $\{A, T, C, G\}$, generated using Matlab's \texttt{randseq} function, in the array.

\sloppypar Similarly, to select a random subsequence $b_1 \hdots b_m$ from a larger string $a_1 \hdots a_n$,  we first obtain a list $C$ of $m$ distinct random integers ranging from $1$ to $n$. We sort the elements of this list in increasing order, and obtain $C = c_1 \hdots c_m$ where $1 \leq c_i < c_{i+1} \leq n$ for $1 \leq i < m$. We then set $b_i=a_{c_i}$ for $1 \leq i \leq m$. In this manner, we select subsequences of length $m$ from each of the organisms and store them in an array, along with a random sequence of length $m$.

Due to the motivation given earlier that fluctuations in the linearity measures first disappear in samples of length 50,000, we selected substrings and subsequences of length 50,000 from the larger excerpts of genomes.

\subsubsection*{Computation of linearity measures}
In order to calculate the maximum and average deviation from linearity for a given string $S$ of length $m$, we first count the number $x_l$ of letters $l \in \{A, T, C, G\}$ in $S$ (where $\sum x_l = m$). Let $S_t$ be the substring composed of the first $t$ letters in $S$, and $t_l$ be the number of letters $l \in \{A, T, C, G\}$ in the string $S_t$ (where $\sum t_l = t$). We compute the distance $D(S_t)$ from point $(t_A, t_T, t_C, t_G)$ to the line passing through the origin and $(x_A, x_T, x_C, x_G)$ for $1 \leq t \leq m$ with the formula (derived from equation (1))

\begin{eqnarray}
\nonumber D(S_t) = \sqrt{(x_Ak-t_A)^2 + (x_Tk-t_T)^2 + (x_Ck-t_C)^2 + (x_Gk-t_G)^2} \\
\nonumber \mathrm{where} \ \ k=\frac{x_At_A+x_Tt_T+x_Ct_C+x_Gt_G}{x_A^2+x_T^2+x_C^2+x_G^2}.
\end{eqnarray}

We then store the distances $D(S_t)$, $1 \leq t \leq m$ in a list, and find the average and maximum values of this list using elementary techniques, as well as the number of local maxima using the \texttt{peakfinder} function.

\subsection*{Compilation and normalization of data}

After we have obtained an array of $26$ samples (25 biosequences and one random sequence), we calculate the linearity measures $adv$, $mdv$, and $nlm$ for each sample in the array and end up with a $26 \times 3$ array. We repeat this procedure $1,000$ times with different random samples, and attain a 3-dimensional array ($26 \times 3 \times 1,000$). We then take the average over the $1,000$ trials, and again obtain a 2-dimensional array, which we normalize for each linearity measure in order to better see the relationships between organisms. Thus, our final product is a $26 \times 3$ array with values ranging between 0 and 100, which allows us to easily compare organisms based on the three linearity criteria. We repeat the whole procedure twice -- once where the samples are substrings of the larger excerpts of genome, and once where they are subsequences. The results are summarized in Table 1.

\section{Discussion}

In this section we provide further details about our experimental work and discuss the obtained results. We comment on results which are obvious to our unarmed eye, hoping that other interesting conclusions could also be drawn by experts with higher expertise in biological sciences.

\subsection{General observations and comments}

\begin{table}
\renewcommand{\arraystretch}{1.3}

\begin{tabular}{ |l|l||r|r|r||r|r|r| }

\hline
 \multirow{2}{*}{Scientific Name} & \multirow{2}{*}{Common Name} & \multicolumn{3}{|c||}{Substrings} & \multicolumn{3}{|c|}{Subsequences}  \\
\cline{3-8}
& & \multicolumn{1}{|c|}{$adv$} & \multicolumn{1}{|c|}{$mdv$} & \multicolumn{1}{|c||}{$nlm$} & \multicolumn{1}{|c|}{$adv$} & \multicolumn{1}{|c|}{$mdv$} & \multicolumn{1}{|c|}{$nlm$} \\
\hline

Random Sequence	&	Random	&		12.8		&		12.8		&	100.0	&	9.4	&	8.9	&	100.0	\\
H. Sapiens	&	Human	&		88.9		&		89.9		&	14.0	&	84.2	&	87.8	&	7.9	\\
H. Neanderthalensis	&	Neanderthal	&		85.5		&		85.1		&	14.8	&	100.0	&	100.0	&	7.7	\\
G. Gorilla	&	Gorilla	&		100.0		&		100.0		&	16.5	&	61.8	&	69.4	&	4.6	\\
P. Troglodytes	&	Chimp	&		81.2		&		81.0		&	15.8	&	40.5	&	39.3	&	11.8	\\
C. Familiaris	&	Dog	&		71.8		&		74.3		&	18.7	&	37.2	&	33.8	&	17.4	\\
G. Gallus	&	Chicken	&		59.1		&		57.2		&	22.5	&	92.8	&	73.5	&	4.5	\\
C. Jacchus	&	Marmoset	&		55.0		&		55.8		&	24.7	&	14.3	&	13.7	&	56.3	\\
R. Norvegicus	&	Rat	&		68.2		&		68.0		&	22.9	&	31.6	&	30.5	&	13.9	\\
M. Musculus	&	Mouse	&		62.5		&		64.9		&	19.8	&	15.5	&	14.9	&	49.7	\\
O. Anatinus	&	Platypus	&		41.2		&		41.7		&	27.1	&	45.4	&	35.0	&	5.4	\\
A. Carolinensis	&	Lizard	&		31.6		&		32.3		&	39.8	&	11.7	&	11.0	&	67.1	\\
D. Rerio	&	Zebrafish	&		59.6		&		58.5		&	25.6	&	32.9	&	30.1	&	11.1	\\
O. Latipes	&	Medaka fish	&		46.6		&		47.5		&	34.7	&	31.9	&	29.3	&	10.9	\\
D. Melanogaster	&	Fruit fly	&		49.1		&		49.3		&	31.1	&	17.1	&	16.1	&	36.3	\\
C. Intestinalis	&	Sea squirt	&		32.8		&		32.2		&	32.0	&	11.2	&	10.6	&	74.7	\\
C. Elegans	&	Nematode	&		51.6		&		52.4		&	22.7	&	18.5	&	19.8	&	26.3	\\
S. Cerevisiae	&	Yeast	&		39.4		&		39.8		&	27.4	&	12.7	&	11.7	&	74.1	\\
C. Muridarum	&	Chlamydia	&		29.7		&		29.3		&	27.9	&	72.3	&	64.9	&	4.5	\\
M.Tuberculosis	&	Tuberculosis	&		33.8		&		34.0		&	23.2	&	9.7	&	9.3	&	96.2	\\
P. Gingivalis	&	Gingivalis	&		50.0		&		48.1		&	18.5	&	18.7	&	16.9	&	31.9	\\
S. Thermophilus	&	Streptococcus	&		37.2		&		36.5		&	22.6	&	47.8	&	43.1	&	4.5	\\
O. Sativa	&	Rice	&		73.8		&		76.9		&	24.5	&	14.9	&	14.0	&	63.5	\\
Z. Mays 	&	Corn	&		76.1		&		80.3		&	18.9	&	18.5	&	18.3	&	44.5	\\
A. Thaliana	&	Cress	&		44.8		&		44.4		&	27.8	&	11.2	&	10.8	&	78.8	\\
G. Max	&	Soybean	&		52.2		&		53.0		&	22.7	&	23.0	&	18.7	&	25.4	\\

\hline
\hline

\multicolumn{2}{|c||}{Maximum pre-normalized value:} & 554.5 & 1100.8 & 21.9 & 739.5 & 1563.6 & 22.4 \\
\hline

\end{tabular}
\\
\caption{Summary of experimental results}
\end{table}

The first two columns of Table 1 give scientific and common names for the organisms which we have examined. The specific strains of Chlamydia, Tuberculosis, Gingivalis, and Streptococcus are Nigg, CCDC5180, W83, and ND03, respectively. All the DNA we processed was from the first chromosomes of the organisms, except for the fruit fly and the yeast, where the DNA was taken from chromosomes 2L and 4, respectively.

Columns 3-5 of Table 1 show the results of our experiments when the $adv$, $mdv$, and  $nlm$ were measured for substrings taken from the 25 biosequences and one random sequence. The last three columns show the results of computing the measures on subsequences instead of substrings. All measures were computed on samples of length 50,000 and are the average of 1,000 trials. 

Note that for most species, $adv$ and $mdv$ measured from substrings differ by less than 1\%; only for four organisms this difference is more than 2\%, but no more than 4.2\%. When measured from subsequences, the difference between $adv$ and $mdv$ is not always small; although for most species it is less than 2\%, for 6 organisms the difference is more than 4\%, and for a few -- as much as 10\% or 20\%.

This observation is also supported by Figure 2, which shows that the $adv$ graphs are nearly identical in appearance to the $mdv$ graphs -- where the difference is slightly more noticeable in the cases where $adv$ and $mdv$ are computed over subsequences.

We suppose that when subsequences rather than substrings are analyzed, some of the structure of the DNA is lost and the results are not as precise. However, as mentioned above, subsequences are slightly more independent of the length of the sample. When substrings are analyzed the entire structure of the DNA is preserved and taken into consideration, so linearity measures computed over substrings seem to be more informative. 

As can be seen from the fifth and eighth columns of the table, the $nlm$ measure is useful for distinguishing between random sequences and biosequences, but not as good for classifying organisms in terms of biological complexity.

\subsection{Distinction between biosequences and random sequences}

Our first important conclusion is the distinction between the linearity of biosequences and random sequences. All of our experiments show that biosequences have a higher average and maximum deviation from linearity than random sequences. When substrings are analyzed, this difference is very significant. In particular, in Table 1, the normalized $adv$ and $mdv$ for random sequences are both 12.8, whereas the normalized $adv$ and $mdv$ for the organism with the smallest deviation are 29.3 and 29.7, respectively. When subsequences are analyzed, the difference is not as substantial, but random sequences still have a smaller average and maximum deviation from linearity than any of the organisms considered. 

This conclusion is also supported by Figure 2, where in the top four graphs (which are computed over substrings), the line positioned visibly below the others is the one representing the random sequence. In the bottom four graphs (which are computed over subsequences), the lowest line is again the one representing the random sequence, but the difference is not as significant.

The criterion of $nlm$ presents an even more sizeable difference between random sequences and biosequences. Again, the difference is larger when substrings are considered: the normalized $nlm$ for random sequences is 100.0, whereas the normalized $nlm$ for the organism with the largest number of local maxima is 39.8. Computed over subsequences, the normalized $nlm$ for random sequences is again 100.0, and the largest number of local maxima for an organism is 96.2.

\subsection{Gradient between primitive and biologically complex organisms}

Our experiments also support the hypothesis that the sequences of primitive organisms are closer in linearity to random sequences than the sequences of biologically complex organisms. As most primitive organisms, we consider bacteria and microscopic organisms; we consider plants the next most evolved organisms, followed by fish, reptiles, and other egg-laying vertebrates. Finally, we consider mammals and primates as organisms at the top of the evolutionary ladder.  We expected that the graded change in the magnitude of deviation from linearity of different organisms would be in accordance with the aforementioned classification of their biological complexity. Indeed, our experiments support this expectation.

In particular, when measured over substrings, the Human, Neanderthal, Gorilla, and Chimpanzee have the highest $adv$ and $mdv$; the bacterium Chlamydia has the smallest $adv$ and $mdv$ after the random sequence. The other organisms with the lowest deviations from linearity are two other bacteria, the yeast, sea squirt, and lizard. In the mid-low range are organisms like the fruit fly, medaka fish, and soybean plant, and in the mid-high range are organisms like the zebrafish, chicken, and mouse. Another interesting observation is that primates have a lower $nlm$ than other animals. Aside from that, the $nlm$ measure is not as useful as the other measures for classifying organisms by biological complexity; a weak inverse relationship exists between $nlm$ and biological complexity, especially when $nlm$ is measured from substrings.

When subsequences are considered some of the structure of the DNA is inadvertently lost, so the results are not as consistent with expectation. However, many similar trends can be seen in the last three columns of Table 1 as well. Primates generally have the highest deviation from linearity, while bacteria like Tuberculosis and Gingivalis have among the lowest deviations from linearity. Other organisms of medium biological complexity span the intermediate values of the linearity measures. 

In all considerations, some anomalies and incongruences with expectation are manifested. For example, when measured over substrings, the $adv$ and $mdv$ of rice and corn are relatively high -- higher, for example, than the $adv$ and $mdv$ of the mouse and rat. A possible explanation for this is that small rodents are perhaps more ancient organisms than certain plants and have not undergone significant evolution in millions of years.

\section{Concluding remarks}

In this paper we introduced a geometric approach for string analysis based on the notions of string linearity and deviation from linearity.
Our experiments showed that, unlike some other criteria, ours strongly separate random sequences from biosequences,
as well as primitive from biologically complex organisms. 
These results are in accordance with certain earlier interpretations that biosequences have been evolving towards
energy minimization in physical terms, as well as of lowering their information complexity \cite{axa2,pande}.
   
As the proposed quantitative measures seem to be quite robust and reliable in practice, 
important future tasks are seen in performing systematic extensive experiments on a larger set of biosequences,
their interpretation and deeper analysis from a biological point of view, and comparison with results obtained by other approaches.  

In addition to a more extensive study of the general trends exhibited in the present work, 
possible future tasks can pursue understanding the meaning and functions (from a biological point of view)
of biosequence locations where deviation from linearity achieves local maxima or minima.
Certain anomalies from the general trends featured by the experiments could also be addressed -- for example, the higher deviation from linearity of certain plants compared to those of rodents.    

Our study was meant to be a pilot one rather than exhaustive. It was intended to provide initial tests of the proposed approach and to serve as a prelude to a more complete interdisciplinary investigation, involving specialists who are better equipped to carry out large-scale experiments and interpret the results.

As a final remark, we believe that the introduced approach could be applied to other areas where string comparison is relevant --
for instance in comparing the structure of natural languages or diverse encoding schemes.

\end{document}